\documentclass[reprint, secnumarabic,amssymb, aps, prl, superscriptaddress
]{revtex4-2}

\usepackage{graphicx}

\usepackage{epsfig}
\usepackage{amsmath}
\usepackage{hyperref}

\usepackage[UKenglish]{babel} 
\usepackage[separate-uncertainty = true]{siunitx}
\usepackage[utf8]{inputenc}

\begin{document}

\title{
Active reset of a radiative cascade for superequilibrium entangled photon generation
}

\author{Jonathan~R.~A.~Müller}%
\affiliation{Toshiba Research Europe Limited, Cambridge Research Laboratory, 208 Science Park, Milton Road, Cambridge CB4 0GZ, UK}
\affiliation{Department of Physics and Astronomy, University of Sheffield, Hounsfield Road, Sheffield S3 7RH, UK}

\author{R.~Mark~Stevenson}%
\email{mark.stevenson@crl.toshiba.co.uk}
\affiliation{Toshiba Research Europe Limited, Cambridge Research Laboratory, 208 Science Park, Milton Road, Cambridge CB4 0GZ, UK}

\author{Joanna~Skiba-Szymanska}%
\affiliation{Toshiba Research Europe Limited, Cambridge Research Laboratory, 208 Science Park, Milton Road, Cambridge CB4 0GZ, UK}

\author{Ginny~Shooter}%
\affiliation{Toshiba Research Europe Limited, Cambridge Research Laboratory, 208 Science Park, Milton Road, Cambridge CB4 0GZ, UK}
\affiliation{Cavendish Laboratory, University of Cambridge, Madingley Road, Cambridge CB3 0HE, UK}

\author{Jan~Huwer}%
\affiliation{Toshiba Research Europe Limited, Cambridge Research Laboratory, 208 Science Park, Milton Road, Cambridge CB4 0GZ, UK}

\author{Ian~Farrer}%
\altaffiliation[Present address: ]{Department of Electronic and Electrical Engineering, University of Sheffield, Mappin Street, Sheffield S1 3JD, UK}
\affiliation{Cavendish Laboratory, University of Cambridge, Madingley Road, Cambridge CB3 0HE, UK}

\author{David~A.~Ritchie}%
\affiliation{Cavendish Laboratory, University of Cambridge, Madingley Road, Cambridge CB3 0HE, UK}

\author{Andrew~J.~Shields}%
\affiliation{Toshiba Research Europe Limited, Cambridge Research Laboratory, 208 Science Park, Milton Road, Cambridge CB4 0GZ, UK}

\date{\today}%

\begin{abstract}
The generation rate of entangled photons emitted from cascaded few-level systems is intrinsically limited by the lifetime of the radiative transitions.
Here, we overcome this limit for entangled photon pairs from quantum dots via a novel driving regime based on an active reset of the radiative cascade.
We show theoretically and experimentally the driving regime to enable the generation of entangled photon pairs with higher fidelity and intensity compared to the optimum continuously driven equilibrium state. Finally, we electrically generate entangled photon pairs with a total fidelity of \SI{79.5\pm1.1}{\percent} at a record clock rate of 1.15\,GHz.
\end{abstract}

\maketitle

Single and entangled photons promise unique advantages for many applications in quantum photonics, including enhanced secure key rates in quantum key distribution (QKD) through elimination of multiphoton emission \cite{Gisin.2002}, and global scale unconditionally secure networks with entanglement-based quantum repeaters \cite{Dur.1999}.
Semiconductor quantum dots (QDs) have been shown to emit single and entangled photons with approaching ideal properties \cite{Ding.2016, Chen.2018, Hanschke.2018, Huber.2018b, Liu.2019, Wang.2019}, with the prospect of on-chip integration \cite{Shields.2007, Huber.2018}, and compatibility with optical fibre quantum networks \cite{Olbrich.2017, Xiang.2019}.
However, state-of-the-art QKD systems based on coherent laser pulses employ clock rates in the gigahertz range for high data throughput \cite{Yuan.2018, Boaron.2018}. Clock rates achieved with entangled photons from QDs are much lower \cite{Stevenson.2012, Zhang.2015}, intrinsically limited by the lifetime of the transitions \cite{Buckley.2012, Senellart.2017}, impeding high frequency operation.
In this letter, we introduce a novel driving regime based on an active reset of the radiative cascade. By suppressing the ground state population, this scheme allows to exceed the fidelity and emission rate of the optimum continuously driven equilibrium state.
We expect our results will enable improved performance of a wide range of entanglement-based photonic applications.

\begin{figure*}[htp]
  \centering
  \includegraphics[width=0.95\textwidth]{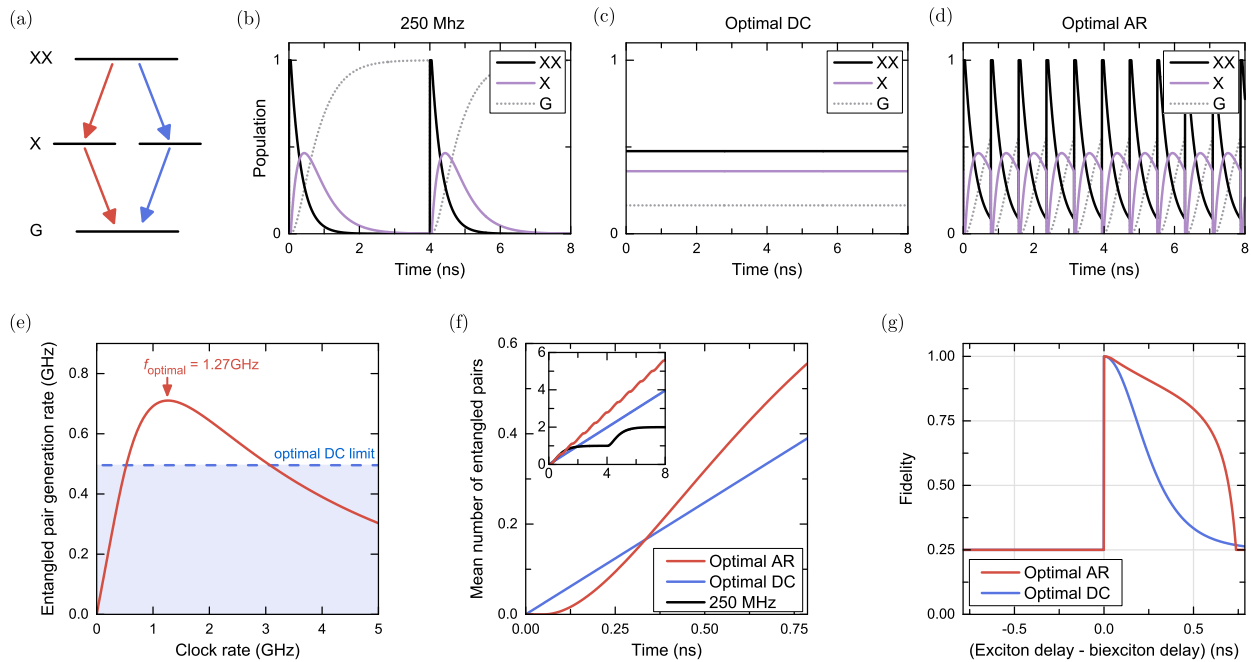}
  \caption{Active reset (AR) of the radiative cascade.
  ({a}) Level diagram of the biexciton-exciton radiative cascade. Due to the superposition of the two separate decay paths (red and blue), the emitted photon pair forms a polarization-entangled state.
  {(b)--(d)} QD populations for different driving modes, calculated via a rate equation model.
  ({b}) Conventional full-cycle pulsed driving. The QD is reinitialised after the population has reached zero.
  ({c}) Conventional DC driving. Populations are shown for the DC pump rate that maximizes the entangled pair emission rate.
  ({d}) Active reset regime. The quantum dot is reinitialised into the biexciton state before the QD population reaches zero.
  ({e}) Modeled time-averaged entangled photon pair generation rate as a function of driving clock rate. An optimum clock rate is given at \SI{1.27}{\giga\hertz}. The pair generation rates accessible in DC driving are indicated by the light-blue-shaded area.
  ({f}), Mean number of entangled photon pairs emitted over one 1.27 GHz period for the driving modes of panels b--d. The inset shows the same for a longer time scale. AR driving (red) generates a higher number of entangled pairs than the optimal DC driving conditions (blue).
  ({g}), Fidelity to a maximally entangled Bell state vs emission delay between exciton and biexciton, for optimal AR driving (red) and optimal DC driving (blue). AR driving maintains a high fidelity even for a maximized pair emission rate.}\label{fig:RR_model_comparison}
\end{figure*}

We begin by considering the excitation and radiative cascade dynamics of an atomic three-level system \cite{*[{for use of the cascade in single atoms see for example }] [{}] Aspect.1981}, as in this case, of a QD. The biexciton state ($XX$) decays via an intermediate neutral exciton ($X$) superposition state, resulting in two consecutively emitted, polarization-entangled photons (Fig.~\ref{fig:RR_model_comparison}a) \cite{Benson.2000, Stevenson.2006}.
We explore this system using a rate equation model
\cite{supplemental}, with the two consecutive transitions assigned radiative lifetimes of $\tau_{XX} = \SI{300}{\pico\second}$ and $\tau_{X} = \SI{500}{\pico\second}$ respectively, similar to observed experimentally.
Fig.~\ref{fig:RR_model_comparison}b shows the modeled  biexciton, exciton, and ground state (G) population, driven at a slow \SI{250}{MHz} clock rate via \SI{50}{\pico\second} rectangular electrical initialization pulses.
When the QD is excited to the $XX$ state in the limit of low driving frequency, practically complete radiative decay of the QD emission is observed
, however a substantial fraction of the clock cycle is spent at the tail of the cascade with high ground state population and low emission brightness.

In contrast, in the limit of high driving frequency, excitation is continuous and equivalent to direct current (DC) driving of the QD (Fig.~\ref{fig:RR_model_comparison}c). Here the QD is in equilibrium, with the $XX$ and $X$ states always populated and an always non-zero emission intensity.
However, at the DC power that maximizes the entangled pair emission rate (here referred to as `optimal DC') the distribution of the QD population in the DC equilibrium means that only a minority (\SI{43}{\percent})
of the emitted photons form part of a biexciton-exciton cascade, resulting in a reduced relative fraction of entangled pair emission.

We introduce a novel active reset (AR) driving scheme based on two core considerations:
Firstly, to actively reset entanglement at a given clock cycle, the QD does not need to return to the ground state. Instead, it is sufficient to reinitialise the QD directly to the biexciton state --- the initial state of the biexciton-exciton cascade. Secondly, resetting the QD before the population has fully cascaded to the ground state allows for an increased entangled-pair emission brightness, as the low-brightness periods in the tail of the emission are eliminated.

Driven in the AR regime at a fast \SI{1.27}{GHz} clock rate (Fig.~\ref{fig:RR_model_comparison}d),
the QD is perpetually kept in an optically active state with high $XX$ or $X$ populations, and low ground state population below 0.56, eliminating dark periods and resulting in a high emission brightness.
For entangled photon pairs
generated from solid state sources, the entanglement fidelity will typically reduce for an increasing emission delay between the two photons, owing to the interaction of the QD with the solid-state environment \cite{Stevenson.2012b}. By reinitialising the QD state early, we avoid the generation of weakly entangled pairs at longer emission delays.

A major consideration for AR driving is the optimum clock rate required to maximize the number of entangled photon pairs over time (Fig.~\ref{fig:RR_model_comparison}e).
Kantner et al.~\cite{Kantner.2017} have shown that for the generation of single photons, an optimum clock rate exists, after which the emission efficiency decreases.
For the case of entangled photon pair generation, at clock periods longer than the lifetimes of both, biexciton and exciton, the entangled-pair generation rate naturally scales approximately linearly with the repetition rate.
At faster clock rates, the pair generation rate in turn reduces due to the non-zero lifetime of the excitonic states, as the delayed emission of the entangled photon pair is interrupted prematurely before the radiative cascade completes. Perhaps surprisingly, the figure shows that there exists a range of frequencies between $\SI{520}{MHz}$ and $\SI{3.07}{GHz}$ for which the entangled pair generation rate exceeds that for continuous driving.
The entangled pair generation rate reaches a maximum at a clock rate of $f_\mathrm{optimal} = \SI{1.27}{GHz}$ where it exceeds the optimum DC pair generation rate by $43\,\%$. Even the highest-clocked experimental demonstrations operate at clock rates below \SI{500}{\mega\hertz}, with none approaching the optimum \cite{Stevenson.2012, Zhang.2015}.

Note that for AR driving, due to the early reinitialization of the QD state, the mean number of emitted photon pairs per clock cycle is less than one (0.56 in this model, Fig.~\ref{fig:RR_model_comparison}f) at $f_\mathrm{optimal}$.
In contrast for the \SI{250}{\mega\hertz} clock rate, (asymptotically) one entangled pair is emitted each clock cycle.
In practice, further external factors affect the overall efficiency as well, such as device electrical bandwidth and collection efficiency.

Lastly, the entanglement fidelity for DC is fundamentally limited by competing uncorrelated emission from continuous excitation (Fig.~\ref{fig:RR_model_comparison}g).
Meanwhile, for AR driving the fidelity is limited only by additional $XX$ emission during the short (\SI{50}{ps}) initialization pulse. 
Altogether, this model demonstrates the feasibility of overcoming the limits on entanglement fidelity and entangled-pair brightness imposed by DC or full-cycle pulsed driving.

\begin{figure*}[t]
  \centering
    \includegraphics[width=\textwidth]{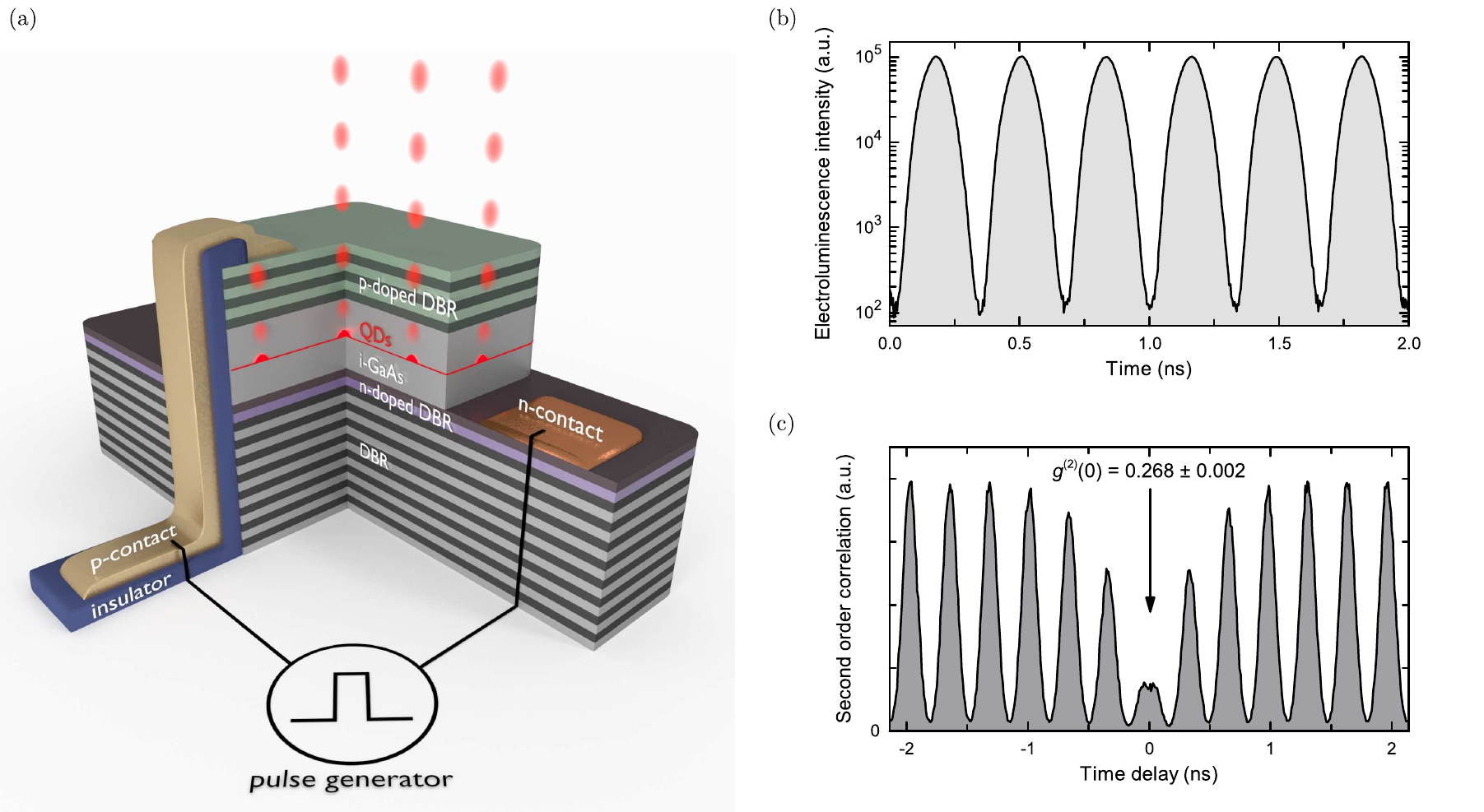}
  \caption{Ultrafast electrical control of an entangled LED. ({a}),~Schematic illustration of the device cross section. Charge carriers are supplied to the QD by inducing an electrical current across the intrinsic region (i-GaAs). A small $\SI{10}{\micro\meter} \times \SI{10}{\micro\meter}$ lateral size of the diode mesa and high-bandwidth optimized packaging (not shown) allow for GHz-clocked operation.
  ({b}),~Time-resolved neutral exciton electroluminescence intensity, driven at \SI{3.05}{\giga\hertz}. 
  ({c}),~Second order correlation $g^{(2)}(t)$ as a function of detection time delay. The emission yields a $g^{(2)}(0)$ value of $0.268 \pm 0.002$ for the integrated center peak. 
  }\label{fig:3ghz_combined}
\end{figure*}
Accessing the regime of gigahertz-clocked generation of single and entangled photons is experimentally challenging. Here we fabricate a high-bandwidth QD LED that responds to very short electrical pulses $<\SI{100}{ps}$, as illustrated in Fig.~\ref{fig:3ghz_combined}a
\cite{supplemental}. A layer of InAs QDs is embedded in a p-i-n diode structure, operated at \SI{6}{\kelvin}.
In order to facilitate gigahertz electrical control of the QD state, we etch the lateral dimensions of the diode mesa to an area of approximately $\SI{10}{\micro\meter} \times \SI{10}{\micro\meter}$, nominally reducing the capacitance to $\sim$\SI{20}{\femto\farad} in the case of a depleted diode.
The n- and p-type contacts are bonded to a high-frequency optimized printed circuit board (PCB) for electrical access.
Sets of 6 (19) AlGaAs-GaAs DBR pairs above (below) the QD layer enhance the collection efficiency of the emitted light.
To verify the high bandwidth of the optimized device design, we investigate the optical signal when applying high frequency voltage pulses.
The device was driven at \SI{3.05}{\giga\hertz} with driving pulses of amplitude $V_{pulse}=\SI{1.43}{\volt}$ and nominal duty cycle of \SI{7.5}{\percent} ($\sim \SI{25}{\pico\second}$).
A DC bias voltage of $V_{DC} =\SI{0.15}{\volt}$ is chosen to be far below the turn-on voltage $V_T=\SI{0.70}{\volt}$ of the diode. Thus, charge carriers quickly tunnel out of the QD potential between driving pulses, quenching the optical emission \cite{Bennett.2005, Hargart.2013} and revealing the electrical bandwidth of the device.

In a time-resolved measurement of the $X$ electroluminescence (Fig.~\ref{fig:3ghz_combined}b) we observe high contrast optical pulses with almost three orders of magnitude modulation of the emission intensity.
Approximating the optical emission pulse with a Gaussian temporal shape yields a FWHM of \SI{100\pm2}{\pico\second}, and fitting an exponential decay yields a lifetime of approximately \SI{15}{\pico\second}. This is in stark contrast to a \SI{500}{\pico\second} radiative lifetime, suggesting non-radiative decay processes are dominant, attributed to efficient tunneling of the charge carriers out of the QD enabled by fast electrical response.

The corresponding second order auto-correlation $g^{(2)}(t)$, measured for the same driving conditions, is shown in Fig.~\ref{fig:3ghz_combined}c. The value $g^{(2)}(0) = {0.268\pm0.002}$ of the zero-delay peak clearly demonstrates the single photon character of the emission. 
We attribute the non-zero component of the $g^{(2)}(0)$ value to background wetting layer emission. The noticeable anti-bunching of the peaks adjacent to the zero-delay could imply the existence of long-lived charged or dark states \cite{Bennett.2005}.
To our knowledge, this is the fastest reported clock rate for the generation of single photons 
\cite{Buckley.2012, Hargart.2013}, entering the ``super high frequency" radio band (comprising the 3--\SI{30}{\giga\hertz} range \cite{InternationalTelecommunicationUnion.2015}) for the first time.

\begin{figure}[tb]
  \centering
  \includegraphics[width=0.45 \textwidth]{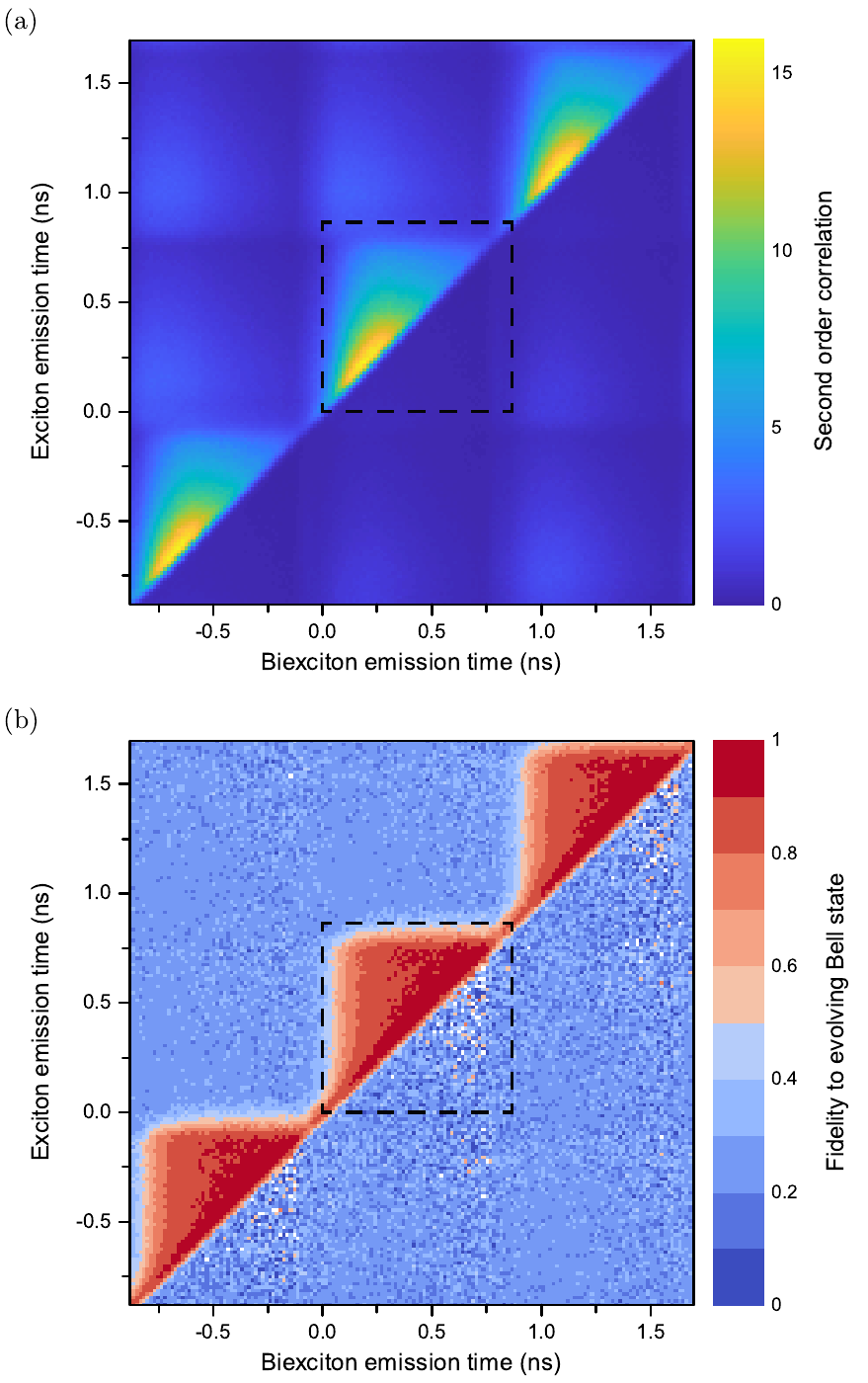}
  \caption{Generation of entangled photon pairs at a 1.15 GHz repetition rate.
The dashed black square highlights photon pairs emitted within the same driving
cycle. The diagonal corresponds to events with simultaneous biexciton and exciton emission time; the measured correlation repeats along the diagonal for integer multiples $n$ of the driving period $t_X, t_{XX} = n \cdot T$.
({a}),~Unpolarized biexciton-exciton second order correlation $g^{(2)}(t_{XX}, t_X)$ as a function of biexciton emission delay $t_{XX}$ and exciton emission delay $t_X$. ({b}),~Corresponding fidelity to a maximally entangled evolving Bell state $f(t_{XX}, t_X)$.
The fidelity remains highly entangled within the driving cycle, and is sharply quenched towards the edge of the cycle $t_X \rightarrow T$. The total fidelity within a cycle is \SI{79.5 \pm 1.1}{\percent}. The data is plotted in bins of $\SI{16}{\pico\second} \times \SI{16}{\pico\second}$; bins with insufficient photon counts are colored in white.}\label{fig:2D_g2_fid}
\end{figure}

After establishing ultrafast control of the QD single photon emission, we now turn our attention towards optimal generation of entangled photon pairs via AR driving, in a proof-of-principle experiment.
The QD LED was electrically excited at a clock rate of \SI{1.15}{\giga\hertz} ($T = \SI{868}{\pico\second}$ period) with initialization pulses of amplitude $V_{pulse}=\SI{0.6}{V}$ and nominal duty cycle \SI{5}{\percent} ($\sim$~\SI{43}{\pico\second}), 
plus a DC bias of $V_{DC} =\SI{0.6}{\volt}$ just below the turn-on voltage. 
Fig.~\ref{fig:2D_g2_fid}a shows the resulting normalized biexciton-exciton second-order correlation function $g^{(2)}(t_{XX}, t_X)$ as a function of $XX$ emission time ($t_{XX}$) and $X$ emission time ($t_X$) relative to the clock of the electrical driving pulse generator.
Due to the cascaded emission of the $X$ photon after the $XX$ photon, the correlation events within a clock cycle (dashed black square) are naturally biased towards the upper triangle region where $t_X > t_{XX}$. The $g^{(2)}(t_{XX}, t_X)$ value remains above 1 for the entire upper triangle of the driving cycle, due to the ground state suppression of AR driving.

Fig.~\ref{fig:2D_g2_fid}b shows the fidelity to a maximally entangled evolving Bell state, extracted from co- and cross-polarized correlations measured in the respective polarization bases \cite{Ward.2014, supplemental}. The emission reaches a maximum fidelity of \SI{95.8\pm1.3}{\percent} and remains above the classical limit of \SI{50}{\percent} for the majority of the cycle, indicating electrically driven photon pair generation at a record \SI{1.15}{GHz} clock rate.
The overall entanglement fidelity, integrated over all photon pairs detected within the same driving cycle (including the `lower triangle' at $t_X < t_{XX}$), comes to $f = \SI{79.5\pm1.1}{\percent}$. For an entanglement-based QKD protocol, this value yields a quantum bit~error~rate of \SI{13.6\pm0.7}{\percent} \cite{Scarani.2009}, well within the \SI{27.6}{\percent} limit required for secure quantum communication \cite{Chau.2002}. Notably, no postselection is required in this driving mode --- weakly entangled photon pairs at longer time delays are avoided naturally, as the entanglement is actively reset at the beginning of each cycle.
Finally, at the edges of the driving cycle, the entanglement is abruptly reduced towards the classically uncorrelated value of \SI{25}{\percent}.
Crucially, the measured overall entanglement fidelity exceeds the overall fidelity of a comparison DC measurement $f_{DC} = \SI{71.2\pm1.0}{\percent}$ for the same time window, taken under the same experimental conditions \cite{supplemental}.

\begin{figure*}[tb]
  \centering
  \includegraphics[width=1 \textwidth]{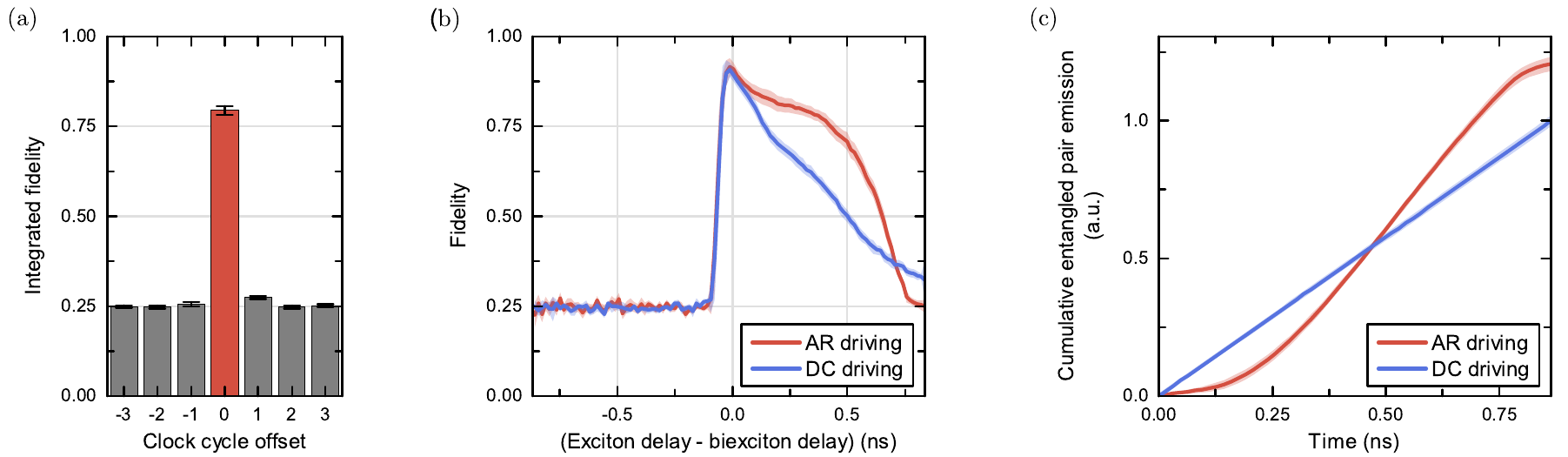}
  \caption{Characteristics of active reset driving at \SI{1.15}{\giga\hertz}. ({a})~Reliable reset of entanglement. Mean entanglement fidelity as a function of clock cycles between biexciton photon detection and exciton photon detection. The mean entanglement is non-classical only for photon pairs detected within the same clock cycle.
  ({b})~Comparison of the entanglement fidelity for AR and DC driving. Histogram of the measured entanglement fidelity as a function of the exciton photon detection delay after the biexciton photon, $t_{X-XX} = t_X - t_{XX}$. The fidelity in AR driving, for coincidences emitted within the same cycle (red line), remains substantially higher than in DC driving (blue line) for the majority of the cycle.
  ({c})~Cumulative entangled pair emission. Plotted over one 1.15GHz cycle, normalized by the overall two-photon coincidences in the respective experiment. After one clock cycle, AR driving yields $(21\pm3)\,\%$ more entangled pairs. Error bars and shaded areas indicate the respective estimated standard error based on Poissonian detection statistics and estimated setup drifts in polarization.}\label{fig:fidelity_combined}
\end{figure*}

An essential factor for the performance of QKD systems
are independent polarizations for photons emitted across subsequent clock cycles, in order to maintain a low quantum bit error rate~\cite{Scarani.2009}. Fig.~\ref{fig:fidelity_combined}a shows the mean entanglement fidelity as a function of the number of clock cycles between $XX$ photon and $X$ photon detection. Importantly, only photon pairs emitted within the same clock cycle are entangled. Photon pairs emitted across different clock cycles in turn yield a mean fidelity close to the \SI{25}{\percent} mark of fully uncorrelated light. This implies that the QD state is reset efficiently at the beginning of each cycle, such that the entanglement does not carry over from one clock cycle to the next. Though reinitialisation to the $XX$ state is dominant, other unentangled initial states may be formed due to the statistical nature of non-resonant excitation such as charged excitons and biexcitons. 

For pairs detected within the same clock cycle, the measured fidelity as a function of time delay between the two photons
(shown in Fig.~\ref{fig:fidelity_combined}b) resembles the concave shape predicted via the rate equation model in Fig.~\ref{fig:RR_model_comparison}g.
For time delays approaching the end of the \SI{868}{\pico\second} repetition period, entanglement is quickly quenched.
Remarkably, as predicted the measured fidelity in AR driving remains non-classical for longer time delays than in DC driving.
At the same time, the fidelity in DC driving remains significantly above \SI{25}{\percent} at the end of the repetition period, thus carrying over an undesirable polarization correlation into the next clock cycle.

Finally, we compare the measured entangled photon pair intensity in AR and DC, accumulated as a function of time delay relative to the clock signal \cite{supplemental}. The form of the curves shown in Fig.~\ref{fig:fidelity_combined}c resemble those in Fig.~\ref{fig:RR_model_comparison}f, with constant pair emission rate for DC driving, and for AR the rate is highest during the central part of the clock period. The measurements show an entangled photon pair rate enhanced by $(21\pm3)\,\%$ for \SI{1.15}{GHz} AR driving compared to continuous driving in DC. This is of similar order to the predicted \SI{43}{\percent} seen in Fig.~\ref{fig:RR_model_comparison}f, with differences attributed to more complex dynamics in the experiment compared to the simple illustrative model (see Supplementary information). However, the implications of this result are clear; resetting the biexciton state before the system is likely in the ground state achieves a higher entangled photon emission rate than sustaining a continuously driven equilibrium.

The super-high-frequency driven (\SI{3.05}{GHz}) single photon LED and active reset driven (1.15GHz) entangled LED results presented above impact other quantum emitter technology, particularly those based on QDs. For electrically driven devices, a key achievement is the high bandwidth entangled-LED mesa structure itself, which places no particular constraints on the optical device design or collection optics. The approach is therefore compatible with a wide variety of other techniques to enhance the source brightness, such as micropillars \cite{Dousse.2010, Ding.2016}, broadband antennas \cite{Chen.2018}, circular Bragg grating cavities \cite{Liu.2019, Wang.2019}, or photonic crystals \cite{Arcari.2014}. Notably, this driving scheme could benefit from Purcell-enhanced radiative decay rates some of these approaches provide, reducing the radiative lifetime and shifting the optimum AR clock rate to even higher frequencies.

Entangled LEDs operated in the active reset regime may benefit the overall performance of entanglement-based photonic applications. The GHz-clocked generation of entangled photon pairs combined with the enhanced entanglement fidelity and source brightness compared to equilibrium operation is of particular interest for entanglement-based QKD protocols, quantum relays, and future implementations of a quantum repeater.


The authors thank Marco Lucamarini and Matthew Anderson for helpful discussions.
This project has received partial funding from the European Union’s Horizon 2020 research and innovation programme under the Marie Skłodowska-Curie grant agreement No 721394 and the Innovate UK project FQNet. JRAM thanks A.~Tartakovskii for academic supervision.

%
%


\begin{thebibliography}{32}%
\makeatletter
\providecommand \@ifxundefined [1]{%
 \@ifx{#1\undefined}
}%
\providecommand \@ifnum [1]{%
 \ifnum #1\expandafter \@firstoftwo
 \else \expandafter \@secondoftwo
 \fi
}%
\providecommand \@ifx [1]{%
 \ifx #1\expandafter \@firstoftwo
 \else \expandafter \@secondoftwo
 \fi
}%
\providecommand \natexlab [1]{#1}%
\providecommand \enquote  [1]{``#1''}%
\providecommand \bibnamefont  [1]{#1}%
\providecommand \bibfnamefont [1]{#1}%
\providecommand \citenamefont [1]{#1}%
\providecommand \href@noop [0]{\@secondoftwo}%
\providecommand \href [0]{\begingroup \@sanitize@url \@href}%
\providecommand \@href[1]{\@@startlink{#1}\@@href}%
\providecommand \@@href[1]{\endgroup#1\@@endlink}%
\providecommand \@sanitize@url [0]{\catcode `\\12\catcode `\$12\catcode
  `\&12\catcode `\#12\catcode `\^12\catcode `\_12\catcode `\%12\relax}%
\providecommand \@@startlink[1]{}%
\providecommand \@@endlink[0]{}%
\providecommand \url  [0]{\begingroup\@sanitize@url \@url }%
\providecommand \@url [1]{\endgroup\@href {#1}{\urlprefix }}%
\providecommand \urlprefix  [0]{URL }%
\providecommand \Eprint [0]{\href }%
\providecommand \doibase [0]{https://doi.org/}%
\providecommand \selectlanguage [0]{\@gobble}%
\providecommand \bibinfo  [0]{\@secondoftwo}%
\providecommand \bibfield  [0]{\@secondoftwo}%
\providecommand \translation [1]{[#1]}%
\providecommand \BibitemOpen [0]{}%
\providecommand \bibitemStop [0]{}%
\providecommand \bibitemNoStop [0]{.\EOS\space}%
\providecommand \EOS [0]{\spacefactor3000\relax}%
\providecommand \BibitemShut  [1]{\csname bibitem#1\endcsname}%
\let\auto@bib@innerbib\@empty
\bibitem [{\citenamefont {Gisin}\ \emph {et~al.}(2002)\citenamefont {Gisin},
  \citenamefont {Ribordy}, \citenamefont {Tittel},\ and\ \citenamefont
  {Zbinden}}]{Gisin.2002}%
  \BibitemOpen
  \bibfield  {author} {\bibinfo {author} {\bibfnamefont {N.}~\bibnamefont
  {Gisin}}, \bibinfo {author} {\bibfnamefont {G.}~\bibnamefont {Ribordy}},
  \bibinfo {author} {\bibfnamefont {W.}~\bibnamefont {Tittel}},\ and\ \bibinfo
  {author} {\bibfnamefont {H.}~\bibnamefont {Zbinden}},\ }\href
  {https://doi.org/10.1103/RevModPhys.74.145} {\bibfield  {journal} {\bibinfo
  {journal} {Rev. Mod. Phys.}\ }\textbf {\bibinfo {volume} {74}},\ \bibinfo
  {pages} {145} (\bibinfo {year} {2002})}\BibitemShut {NoStop}%
\bibitem [{\citenamefont {Dür}\ \emph {et~al.}(1999)\citenamefont {Dür},
  \citenamefont {Briegel}, \citenamefont {Cirac},\ and\ \citenamefont
  {Zoller}}]{Dur.1999}%
  \BibitemOpen
  \bibfield  {author} {\bibinfo {author} {\bibfnamefont {W.}~\bibnamefont
  {Dür}}, \bibinfo {author} {\bibfnamefont {H.-J.}\ \bibnamefont {Briegel}},
  \bibinfo {author} {\bibfnamefont {J.~I.}\ \bibnamefont {Cirac}},\ and\
  \bibinfo {author} {\bibfnamefont {P.}~\bibnamefont {Zoller}},\ }\href
  {https://doi.org/10.1103/PhysRevA.59.169} {\bibfield  {journal} {\bibinfo
  {journal} {Phys. Rev. A}\ }\textbf {\bibinfo {volume} {59}},\ \bibinfo
  {pages} {169} (\bibinfo {year} {1999})}\BibitemShut {NoStop}%
\bibitem [{\citenamefont {Ding}\ \emph {et~al.}(2016)\citenamefont {Ding},
  \citenamefont {He}, \citenamefont {Duan}, \citenamefont {Gregersen},
  \citenamefont {Chen}, \citenamefont {Unsleber}, \citenamefont {Maier},
  \citenamefont {Schneider}, \citenamefont {Kamp}, \citenamefont {Höfling},
  \citenamefont {Lu},\ and\ \citenamefont {Pan}}]{Ding.2016}%
  \BibitemOpen
  \bibfield  {author} {\bibinfo {author} {\bibfnamefont {X.}~\bibnamefont
  {Ding}}, \bibinfo {author} {\bibfnamefont {Y.}~\bibnamefont {He}}, \bibinfo
  {author} {\bibfnamefont {Z.-C.}\ \bibnamefont {Duan}}, \bibinfo {author}
  {\bibfnamefont {N.}~\bibnamefont {Gregersen}}, \bibinfo {author}
  {\bibfnamefont {M.-C.}\ \bibnamefont {Chen}}, \bibinfo {author}
  {\bibfnamefont {S.}~\bibnamefont {Unsleber}}, \bibinfo {author}
  {\bibfnamefont {S.}~\bibnamefont {Maier}}, \bibinfo {author} {\bibfnamefont
  {C.}~\bibnamefont {Schneider}}, \bibinfo {author} {\bibfnamefont
  {M.}~\bibnamefont {Kamp}}, \bibinfo {author} {\bibfnamefont {S.}~\bibnamefont
  {Höfling}}, \bibinfo {author} {\bibfnamefont {C.-Y.}\ \bibnamefont {Lu}},\
  and\ \bibinfo {author} {\bibfnamefont {J.-W.}\ \bibnamefont {Pan}},\ }\href
  {https://doi.org/10.1103/PhysRevLett.116.020401} {\bibfield  {journal}
  {\bibinfo  {journal} {Phys. Rev. Lett.}\ }\textbf {\bibinfo {volume} {116}},\
  \bibinfo {pages} {020401} (\bibinfo {year} {2016})}\BibitemShut {NoStop}%
\bibitem [{\citenamefont {Chen}\ \emph {et~al.}(2018)\citenamefont {Chen},
  \citenamefont {Zopf}, \citenamefont {Keil}, \citenamefont {Ding},\ and\
  \citenamefont {Schmidt}}]{Chen.2018}%
  \BibitemOpen
  \bibfield  {author} {\bibinfo {author} {\bibfnamefont {Y.}~\bibnamefont
  {Chen}}, \bibinfo {author} {\bibfnamefont {M.}~\bibnamefont {Zopf}}, \bibinfo
  {author} {\bibfnamefont {R.}~\bibnamefont {Keil}}, \bibinfo {author}
  {\bibfnamefont {F.}~\bibnamefont {Ding}},\ and\ \bibinfo {author}
  {\bibfnamefont {O.~G.}\ \bibnamefont {Schmidt}},\ }\href
  {https://doi.org/10.1038/s41467-018-05456-2} {\bibfield  {journal} {\bibinfo
  {journal} {Nat Commun}\ }\textbf {\bibinfo {volume} {9}},\ \bibinfo {pages}
  {2994} (\bibinfo {year} {2018})}\BibitemShut {NoStop}%
\bibitem [{\citenamefont {Hanschke}\ \emph {et~al.}(2018)\citenamefont
  {Hanschke}, \citenamefont {Fischer}, \citenamefont {Appel}, \citenamefont
  {Lukin}, \citenamefont {Wierzbowski}, \citenamefont {Sun}, \citenamefont
  {Trivedi}, \citenamefont {Vučković}, \citenamefont {Finley},\ and\
  \citenamefont {Müller}}]{Hanschke.2018}%
  \BibitemOpen
  \bibfield  {author} {\bibinfo {author} {\bibfnamefont {L.}~\bibnamefont
  {Hanschke}}, \bibinfo {author} {\bibfnamefont {K.~A.}\ \bibnamefont
  {Fischer}}, \bibinfo {author} {\bibfnamefont {S.}~\bibnamefont {Appel}},
  \bibinfo {author} {\bibfnamefont {D.}~\bibnamefont {Lukin}}, \bibinfo
  {author} {\bibfnamefont {J.}~\bibnamefont {Wierzbowski}}, \bibinfo {author}
  {\bibfnamefont {S.}~\bibnamefont {Sun}}, \bibinfo {author} {\bibfnamefont
  {R.}~\bibnamefont {Trivedi}}, \bibinfo {author} {\bibfnamefont
  {J.}~\bibnamefont {Vučković}}, \bibinfo {author} {\bibfnamefont {J.~J.}\
  \bibnamefont {Finley}},\ and\ \bibinfo {author} {\bibfnamefont
  {K.}~\bibnamefont {Müller}},\ }\href
  {https://doi.org/10.1038/s41534-018-0092-0} {\bibfield  {journal} {\bibinfo
  {journal} {npj Quantum Inf}\ }\textbf {\bibinfo {volume} {4}},\ \bibinfo
  {pages} {1026} (\bibinfo {year} {2018})}\BibitemShut {NoStop}%
\bibitem [{\citenamefont {Huber}\ \emph
  {et~al.}(2018{\natexlab{a}})\citenamefont {Huber}, \citenamefont {Reindl},
  \citenamefont {{Covre da Silva}}, \citenamefont {Schimpf}, \citenamefont
  {Martín-Sánchez}, \citenamefont {Huang}, \citenamefont {Piredda},
  \citenamefont {Edlinger}, \citenamefont {Rastelli},\ and\ \citenamefont
  {Trotta}}]{Huber.2018b}%
  \BibitemOpen
  \bibfield  {author} {\bibinfo {author} {\bibfnamefont {D.}~\bibnamefont
  {Huber}}, \bibinfo {author} {\bibfnamefont {M.}~\bibnamefont {Reindl}},
  \bibinfo {author} {\bibfnamefont {S.~F.}\ \bibnamefont {{Covre da Silva}}},
  \bibinfo {author} {\bibfnamefont {C.}~\bibnamefont {Schimpf}}, \bibinfo
  {author} {\bibfnamefont {J.}~\bibnamefont {Martín-Sánchez}}, \bibinfo
  {author} {\bibfnamefont {H.}~\bibnamefont {Huang}}, \bibinfo {author}
  {\bibfnamefont {G.}~\bibnamefont {Piredda}}, \bibinfo {author} {\bibfnamefont
  {J.}~\bibnamefont {Edlinger}}, \bibinfo {author} {\bibfnamefont
  {A.}~\bibnamefont {Rastelli}},\ and\ \bibinfo {author} {\bibfnamefont
  {R.}~\bibnamefont {Trotta}},\ }\href
  {https://doi.org/10.1103/PhysRevLett.121.033902} {\bibfield  {journal}
  {\bibinfo  {journal} {Phys. Rev. Lett.}\ }\textbf {\bibinfo {volume} {121}},\
  \bibinfo {pages} {033902} (\bibinfo {year} {2018}{\natexlab{a}})}\BibitemShut
  {NoStop}%
\bibitem [{\citenamefont {Liu}\ \emph {et~al.}(2019)\citenamefont {Liu},
  \citenamefont {Su}, \citenamefont {Wei}, \citenamefont {Yao}, \citenamefont
  {Silva}, \citenamefont {Yu}, \citenamefont {Iles-Smith}, \citenamefont
  {Srinivasan}, \citenamefont {Rastelli}, \citenamefont {Li},\ and\
  \citenamefont {Wang}}]{Liu.2019}%
  \BibitemOpen
  \bibfield  {author} {\bibinfo {author} {\bibfnamefont {J.}~\bibnamefont
  {Liu}}, \bibinfo {author} {\bibfnamefont {R.}~\bibnamefont {Su}}, \bibinfo
  {author} {\bibfnamefont {Y.}~\bibnamefont {Wei}}, \bibinfo {author}
  {\bibfnamefont {B.}~\bibnamefont {Yao}}, \bibinfo {author} {\bibfnamefont
  {S.~F. C.~d.}\ \bibnamefont {Silva}}, \bibinfo {author} {\bibfnamefont
  {Y.}~\bibnamefont {Yu}}, \bibinfo {author} {\bibfnamefont {J.}~\bibnamefont
  {Iles-Smith}}, \bibinfo {author} {\bibfnamefont {K.}~\bibnamefont
  {Srinivasan}}, \bibinfo {author} {\bibfnamefont {A.}~\bibnamefont
  {Rastelli}}, \bibinfo {author} {\bibfnamefont {J.}~\bibnamefont {Li}},\ and\
  \bibinfo {author} {\bibfnamefont {X.}~\bibnamefont {Wang}},\ }\href
  {https://doi.org/10.1038/s41565-019-0435-9} {\bibfield  {journal} {\bibinfo
  {journal} {Nat Nanotech}\ }\textbf {\bibinfo {volume} {14}},\ \bibinfo
  {pages} {586} (\bibinfo {year} {2019})}\BibitemShut {NoStop}%
\bibitem [{\citenamefont {Wang}\ \emph {et~al.}(2019)\citenamefont {Wang},
  \citenamefont {Hu}, \citenamefont {Chung}, \citenamefont {Qin}, \citenamefont
  {Yang}, \citenamefont {Li}, \citenamefont {Liu}, \citenamefont {Zhong},
  \citenamefont {He}, \citenamefont {Ding}, \citenamefont {Deng}, \citenamefont
  {Dai}, \citenamefont {Huo}, \citenamefont {Höfling}, \citenamefont {Lu},\
  and\ \citenamefont {Pan}}]{Wang.2019}%
  \BibitemOpen
  \bibfield  {author} {\bibinfo {author} {\bibfnamefont {H.}~\bibnamefont
  {Wang}}, \bibinfo {author} {\bibfnamefont {H.}~\bibnamefont {Hu}}, \bibinfo
  {author} {\bibfnamefont {T.-H.}\ \bibnamefont {Chung}}, \bibinfo {author}
  {\bibfnamefont {J.}~\bibnamefont {Qin}}, \bibinfo {author} {\bibfnamefont
  {X.}~\bibnamefont {Yang}}, \bibinfo {author} {\bibfnamefont {J.-P.}\
  \bibnamefont {Li}}, \bibinfo {author} {\bibfnamefont {R.-Z.}\ \bibnamefont
  {Liu}}, \bibinfo {author} {\bibfnamefont {H.-S.}\ \bibnamefont {Zhong}},
  \bibinfo {author} {\bibfnamefont {Y.-M.}\ \bibnamefont {He}}, \bibinfo
  {author} {\bibfnamefont {X.}~\bibnamefont {Ding}}, \bibinfo {author}
  {\bibfnamefont {Y.-H.}\ \bibnamefont {Deng}}, \bibinfo {author}
  {\bibfnamefont {Q.}~\bibnamefont {Dai}}, \bibinfo {author} {\bibfnamefont
  {Y.-H.}\ \bibnamefont {Huo}}, \bibinfo {author} {\bibfnamefont
  {S.}~\bibnamefont {Höfling}}, \bibinfo {author} {\bibfnamefont {C.-Y.}\
  \bibnamefont {Lu}},\ and\ \bibinfo {author} {\bibfnamefont {J.-W.}\
  \bibnamefont {Pan}},\ }\href {https://doi.org/10.1103/PhysRevLett.122.113602}
  {\bibfield  {journal} {\bibinfo  {journal} {Phys. Rev. Lett.}\ }\textbf
  {\bibinfo {volume} {122}},\ \bibinfo {pages} {113602} (\bibinfo {year}
  {2019})}\BibitemShut {NoStop}%
\bibitem [{\citenamefont {Shields}(2007)}]{Shields.2007}%
  \BibitemOpen
  \bibfield  {author} {\bibinfo {author} {\bibfnamefont {A.~J.}\ \bibnamefont
  {Shields}},\ }\href {https://doi.org/10.1038/nphoton.2007.46} {\bibfield
  {journal} {\bibinfo  {journal} {Nature Photon}\ }\textbf {\bibinfo {volume}
  {1}},\ \bibinfo {pages} {215} (\bibinfo {year} {2007})}\BibitemShut {NoStop}%
\bibitem [{\citenamefont {Huber}\ \emph
  {et~al.}(2018{\natexlab{b}})\citenamefont {Huber}, \citenamefont {Reindl},
  \citenamefont {Aberl}, \citenamefont {Rastelli},\ and\ \citenamefont
  {Trotta}}]{Huber.2018}%
  \BibitemOpen
  \bibfield  {author} {\bibinfo {author} {\bibfnamefont {D.}~\bibnamefont
  {Huber}}, \bibinfo {author} {\bibfnamefont {M.}~\bibnamefont {Reindl}},
  \bibinfo {author} {\bibfnamefont {J.}~\bibnamefont {Aberl}}, \bibinfo
  {author} {\bibfnamefont {A.}~\bibnamefont {Rastelli}},\ and\ \bibinfo
  {author} {\bibfnamefont {R.}~\bibnamefont {Trotta}},\ }\href
  {https://doi.org/10.1088/2040-8986/aac4c4} {\bibfield  {journal} {\bibinfo
  {journal} {J. Opt.}\ }\textbf {\bibinfo {volume} {20}},\ \bibinfo {pages}
  {073002} (\bibinfo {year} {2018}{\natexlab{b}})}\BibitemShut {NoStop}%
\bibitem [{\citenamefont {Olbrich}\ \emph {et~al.}(2017)\citenamefont
  {Olbrich}, \citenamefont {Höschele}, \citenamefont {Müller}, \citenamefont
  {Kettler}, \citenamefont {{Luca Portalupi}}, \citenamefont {Paul},
  \citenamefont {Jetter},\ and\ \citenamefont {Michler}}]{Olbrich.2017}%
  \BibitemOpen
  \bibfield  {author} {\bibinfo {author} {\bibfnamefont {F.}~\bibnamefont
  {Olbrich}}, \bibinfo {author} {\bibfnamefont {J.}~\bibnamefont {Höschele}},
  \bibinfo {author} {\bibfnamefont {M.}~\bibnamefont {Müller}}, \bibinfo
  {author} {\bibfnamefont {J.}~\bibnamefont {Kettler}}, \bibinfo {author}
  {\bibfnamefont {S.}~\bibnamefont {{Luca Portalupi}}}, \bibinfo {author}
  {\bibfnamefont {M.}~\bibnamefont {Paul}}, \bibinfo {author} {\bibfnamefont
  {M.}~\bibnamefont {Jetter}},\ and\ \bibinfo {author} {\bibfnamefont
  {P.}~\bibnamefont {Michler}},\ }\href {https://doi.org/10.1063/1.4994145}
  {\bibfield  {journal} {\bibinfo  {journal} {Appl. Phys. Lett.}\ }\textbf
  {\bibinfo {volume} {111}},\ \bibinfo {pages} {133106} (\bibinfo {year}
  {2017})}\BibitemShut {NoStop}%
\bibitem [{\citenamefont {Xiang}\ \emph {et~al.}(2019)\citenamefont {Xiang},
  \citenamefont {Huwer}, \citenamefont {Stevenson}, \citenamefont
  {Skiba-Szymanska}, \citenamefont {Ward}, \citenamefont {Farrer},
  \citenamefont {Ritchie},\ and\ \citenamefont {Shields}}]{Xiang.2019}%
  \BibitemOpen
  \bibfield  {author} {\bibinfo {author} {\bibfnamefont {Z.-H.}\ \bibnamefont
  {Xiang}}, \bibinfo {author} {\bibfnamefont {J.}~\bibnamefont {Huwer}},
  \bibinfo {author} {\bibfnamefont {R.~M.}\ \bibnamefont {Stevenson}}, \bibinfo
  {author} {\bibfnamefont {J.}~\bibnamefont {Skiba-Szymanska}}, \bibinfo
  {author} {\bibfnamefont {M.~B.}\ \bibnamefont {Ward}}, \bibinfo {author}
  {\bibfnamefont {I.}~\bibnamefont {Farrer}}, \bibinfo {author} {\bibfnamefont
  {D.~A.}\ \bibnamefont {Ritchie}},\ and\ \bibinfo {author} {\bibfnamefont
  {A.~J.}\ \bibnamefont {Shields}},\ }\href
  {https://doi.org/10.1038/s41598-019-40912-z} {\bibfield  {journal} {\bibinfo
  {journal} {Sci Rep}\ }\textbf {\bibinfo {volume} {9}},\ \bibinfo {pages}
  {4111} (\bibinfo {year} {2019})}\BibitemShut {NoStop}%
\bibitem [{\citenamefont {Yuan}\ \emph {et~al.}(2018)\citenamefont {Yuan},
  \citenamefont {Plews}, \citenamefont {Takahashi}, \citenamefont {Doi},
  \citenamefont {Tam}, \citenamefont {Sharpe}, \citenamefont {Dixon},
  \citenamefont {Lavelle}, \citenamefont {Dynes}, \citenamefont {Murakami},
  \citenamefont {Kujiraoka}, \citenamefont {Lucamarini}, \citenamefont
  {Tanizawa}, \citenamefont {Sato},\ and\ \citenamefont {Shields}}]{Yuan.2018}%
  \BibitemOpen
  \bibfield  {author} {\bibinfo {author} {\bibfnamefont {Z.}~\bibnamefont
  {Yuan}}, \bibinfo {author} {\bibfnamefont {A.}~\bibnamefont {Plews}},
  \bibinfo {author} {\bibfnamefont {R.}~\bibnamefont {Takahashi}}, \bibinfo
  {author} {\bibfnamefont {K.}~\bibnamefont {Doi}}, \bibinfo {author}
  {\bibfnamefont {W.}~\bibnamefont {Tam}}, \bibinfo {author} {\bibfnamefont
  {A.~W.}\ \bibnamefont {Sharpe}}, \bibinfo {author} {\bibfnamefont {A.~R.}\
  \bibnamefont {Dixon}}, \bibinfo {author} {\bibfnamefont {E.}~\bibnamefont
  {Lavelle}}, \bibinfo {author} {\bibfnamefont {J.~F.}\ \bibnamefont {Dynes}},
  \bibinfo {author} {\bibfnamefont {A.}~\bibnamefont {Murakami}}, \bibinfo
  {author} {\bibfnamefont {M.}~\bibnamefont {Kujiraoka}}, \bibinfo {author}
  {\bibfnamefont {M.}~\bibnamefont {Lucamarini}}, \bibinfo {author}
  {\bibfnamefont {Y.}~\bibnamefont {Tanizawa}}, \bibinfo {author}
  {\bibfnamefont {H.}~\bibnamefont {Sato}},\ and\ \bibinfo {author}
  {\bibfnamefont {A.~J.}\ \bibnamefont {Shields}},\ }\href
  {https://doi.org/10.1109/JLT.2018.2843136} {\bibfield  {journal} {\bibinfo
  {journal} {J. Lightwave Technol.}\ }\textbf {\bibinfo {volume} {36}},\
  \bibinfo {pages} {3427} (\bibinfo {year} {2018})}\BibitemShut {NoStop}%
\bibitem [{\citenamefont {Boaron}\ \emph {et~al.}(2018)\citenamefont {Boaron},
  \citenamefont {Boso}, \citenamefont {Rusca}, \citenamefont {Vulliez},
  \citenamefont {Autebert}, \citenamefont {Caloz}, \citenamefont {Perrenoud},
  \citenamefont {Gras}, \citenamefont {Bussières}, \citenamefont {Li},
  \citenamefont {Nolan}, \citenamefont {Martin},\ and\ \citenamefont
  {Zbinden}}]{Boaron.2018}%
  \BibitemOpen
  \bibfield  {author} {\bibinfo {author} {\bibfnamefont {A.}~\bibnamefont
  {Boaron}}, \bibinfo {author} {\bibfnamefont {G.}~\bibnamefont {Boso}},
  \bibinfo {author} {\bibfnamefont {D.}~\bibnamefont {Rusca}}, \bibinfo
  {author} {\bibfnamefont {C.}~\bibnamefont {Vulliez}}, \bibinfo {author}
  {\bibfnamefont {C.}~\bibnamefont {Autebert}}, \bibinfo {author}
  {\bibfnamefont {M.}~\bibnamefont {Caloz}}, \bibinfo {author} {\bibfnamefont
  {M.}~\bibnamefont {Perrenoud}}, \bibinfo {author} {\bibfnamefont
  {G.}~\bibnamefont {Gras}}, \bibinfo {author} {\bibfnamefont {F.}~\bibnamefont
  {Bussières}}, \bibinfo {author} {\bibfnamefont {M.-J.}\ \bibnamefont {Li}},
  \bibinfo {author} {\bibfnamefont {D.}~\bibnamefont {Nolan}}, \bibinfo
  {author} {\bibfnamefont {A.}~\bibnamefont {Martin}},\ and\ \bibinfo {author}
  {\bibfnamefont {H.}~\bibnamefont {Zbinden}},\ }\href
  {https://doi.org/10.1103/PhysRevLett.121.190502} {\bibfield  {journal}
  {\bibinfo  {journal} {Phys. Rev. Lett.}\ }\textbf {\bibinfo {volume} {121}},\
  \bibinfo {pages} {190502} (\bibinfo {year} {2018})}\BibitemShut {NoStop}%
\bibitem [{\citenamefont {Stevenson}\ \emph
  {et~al.}(2012{\natexlab{a}})\citenamefont {Stevenson}, \citenamefont
  {Salter}, \citenamefont {Nilsson}, \citenamefont {Bennett}, \citenamefont
  {Ward}, \citenamefont {Farrer}, \citenamefont {Ritchie},\ and\ \citenamefont
  {Shields}}]{Stevenson.2012}%
  \BibitemOpen
  \bibfield  {author} {\bibinfo {author} {\bibfnamefont {R.~M.}\ \bibnamefont
  {Stevenson}}, \bibinfo {author} {\bibfnamefont {C.~L.}\ \bibnamefont
  {Salter}}, \bibinfo {author} {\bibfnamefont {J.}~\bibnamefont {Nilsson}},
  \bibinfo {author} {\bibfnamefont {A.~J.}\ \bibnamefont {Bennett}}, \bibinfo
  {author} {\bibfnamefont {M.~B.}\ \bibnamefont {Ward}}, \bibinfo {author}
  {\bibfnamefont {I.}~\bibnamefont {Farrer}}, \bibinfo {author} {\bibfnamefont
  {D.~A.}\ \bibnamefont {Ritchie}},\ and\ \bibinfo {author} {\bibfnamefont
  {A.~J.}\ \bibnamefont {Shields}},\ }\href
  {https://doi.org/10.1103/PhysRevLett.108.040503} {\bibfield  {journal}
  {\bibinfo  {journal} {Phys. Rev. Lett.}\ }\textbf {\bibinfo {volume} {108}},\
  \bibinfo {pages} {040503} (\bibinfo {year} {2012}{\natexlab{a}})}\BibitemShut
  {NoStop}%
\bibitem [{\citenamefont {Zhang}\ \emph {et~al.}(2015)\citenamefont {Zhang},
  \citenamefont {Wildmann}, \citenamefont {Ding}, \citenamefont {Trotta},
  \citenamefont {Huo}, \citenamefont {Zallo}, \citenamefont {Huber},
  \citenamefont {Rastelli},\ and\ \citenamefont {Schmidt}}]{Zhang.2015}%
  \BibitemOpen
  \bibfield  {author} {\bibinfo {author} {\bibfnamefont {J.}~\bibnamefont
  {Zhang}}, \bibinfo {author} {\bibfnamefont {J.~S.}\ \bibnamefont {Wildmann}},
  \bibinfo {author} {\bibfnamefont {F.}~\bibnamefont {Ding}}, \bibinfo {author}
  {\bibfnamefont {R.}~\bibnamefont {Trotta}}, \bibinfo {author} {\bibfnamefont
  {Y.}~\bibnamefont {Huo}}, \bibinfo {author} {\bibfnamefont {E.}~\bibnamefont
  {Zallo}}, \bibinfo {author} {\bibfnamefont {D.}~\bibnamefont {Huber}},
  \bibinfo {author} {\bibfnamefont {A.}~\bibnamefont {Rastelli}},\ and\
  \bibinfo {author} {\bibfnamefont {O.~G.}\ \bibnamefont {Schmidt}},\ }\href
  {https://doi.org/10.1038/ncomms10067} {\bibfield  {journal} {\bibinfo
  {journal} {Nat Commun}\ }\textbf {\bibinfo {volume} {6}},\ \bibinfo {pages}
  {10067} (\bibinfo {year} {2015})}\BibitemShut {NoStop}%
\bibitem [{\citenamefont {Buckley}\ \emph {et~al.}(2012)\citenamefont
  {Buckley}, \citenamefont {Rivoire},\ and\ \citenamefont
  {Vučković}}]{Buckley.2012}%
  \BibitemOpen
  \bibfield  {author} {\bibinfo {author} {\bibfnamefont {S.}~\bibnamefont
  {Buckley}}, \bibinfo {author} {\bibfnamefont {K.}~\bibnamefont {Rivoire}},\
  and\ \bibinfo {author} {\bibfnamefont {J.}~\bibnamefont {Vučković}},\
  }\href {https://doi.org/10.1088/0034-4885/75/12/126503} {\bibfield  {journal}
  {\bibinfo  {journal} {Rep Prog Phys}\ }\textbf {\bibinfo {volume} {75}},\
  \bibinfo {pages} {126503} (\bibinfo {year} {2012})}\BibitemShut {NoStop}%
\bibitem [{\citenamefont {Senellart}\ \emph {et~al.}(2017)\citenamefont
  {Senellart}, \citenamefont {Solomon},\ and\ \citenamefont
  {White}}]{Senellart.2017}%
  \BibitemOpen
  \bibfield  {author} {\bibinfo {author} {\bibfnamefont {P.}~\bibnamefont
  {Senellart}}, \bibinfo {author} {\bibfnamefont {G.}~\bibnamefont {Solomon}},\
  and\ \bibinfo {author} {\bibfnamefont {A.}~\bibnamefont {White}},\ }\href
  {https://doi.org/10.1038/nnano.2017.218} {\bibfield  {journal} {\bibinfo
  {journal} {Nat Nanotech}\ }\textbf {\bibinfo {volume} {12}},\ \bibinfo
  {pages} {1026} (\bibinfo {year} {2017})}\BibitemShut {NoStop}%
\bibitem [{\citenamefont {Aspect}\ \emph {et~al.}(1981)\citenamefont {Aspect},
  \citenamefont {Grangier},\ and\ \citenamefont {Roger}}]{Aspect.1981}%
  \BibitemOpen
  \bibfield  {author} {\bibinfo {author} {\bibfnamefont {A.}~\bibnamefont
  {Aspect}}, \bibinfo {author} {\bibfnamefont {P.}~\bibnamefont {Grangier}},\
  and\ \bibinfo {author} {\bibfnamefont {G.}~\bibnamefont {Roger}},\ }\href
  {https://doi.org/10.1103/PhysRevLett.47.460} {\bibfield  {journal} {\bibinfo
  {journal} {Phys. Rev. Lett.}\ }\textbf {\bibinfo {volume} {47}},\ \bibinfo
  {pages} {460} (\bibinfo {year} {1981})}\BibitemShut {NoStop}%
\bibitem [{\citenamefont {Benson}\ \emph {et~al.}(2000)\citenamefont {Benson},
  \citenamefont {Santori}, \citenamefont {Pelton},\ and\ \citenamefont
  {Yamamoto}}]{Benson.2000}%
  \BibitemOpen
  \bibfield  {author} {\bibinfo {author} {\bibfnamefont {O.}~\bibnamefont
  {Benson}}, \bibinfo {author} {\bibfnamefont {C.}~\bibnamefont {Santori}},
  \bibinfo {author} {\bibfnamefont {M.}~\bibnamefont {Pelton}},\ and\ \bibinfo
  {author} {\bibfnamefont {Y.}~\bibnamefont {Yamamoto}},\ }\href
  {https://doi.org/10.1103/PhysRevLett.84.2513} {\bibfield  {journal} {\bibinfo
   {journal} {Phys. Rev. Lett.}\ }\textbf {\bibinfo {volume} {84}},\ \bibinfo
  {pages} {2513} (\bibinfo {year} {2000})}\BibitemShut {NoStop}%
\bibitem [{\citenamefont {Stevenson}\ \emph {et~al.}(2006)\citenamefont
  {Stevenson}, \citenamefont {Young}, \citenamefont {Atkinson}, \citenamefont
  {Cooper}, \citenamefont {Ritchie},\ and\ \citenamefont
  {Shields}}]{Stevenson.2006}%
  \BibitemOpen
  \bibfield  {author} {\bibinfo {author} {\bibfnamefont {R.~M.}\ \bibnamefont
  {Stevenson}}, \bibinfo {author} {\bibfnamefont {R.~J.}\ \bibnamefont
  {Young}}, \bibinfo {author} {\bibfnamefont {P.}~\bibnamefont {Atkinson}},
  \bibinfo {author} {\bibfnamefont {K.}~\bibnamefont {Cooper}}, \bibinfo
  {author} {\bibfnamefont {D.~A.}\ \bibnamefont {Ritchie}},\ and\ \bibinfo
  {author} {\bibfnamefont {A.~J.}\ \bibnamefont {Shields}},\ }\href
  {https://doi.org/10.1038/nature04446} {\bibfield  {journal} {\bibinfo
  {journal} {Nature}\ }\textbf {\bibinfo {volume} {439}},\ \bibinfo {pages}
  {179} (\bibinfo {year} {2006})}\BibitemShut {NoStop}%
\bibitem [{sup()}]{supplemental}%
  \BibitemOpen
  \href@noop {} {\bibinfo {title} {{See Supplemental Material at [URL will be
  inserted by publisher] for details on the rate equation model, LED device
  design and experimental configuration}}}\BibitemShut {NoStop}%
\bibitem [{\citenamefont {Stevenson}\ \emph
  {et~al.}(2012{\natexlab{b}})\citenamefont {Stevenson}, \citenamefont
  {Bennett},\ and\ \citenamefont {Shields}}]{Stevenson.2012b}%
  \BibitemOpen
  \bibfield  {author} {\bibinfo {author} {\bibfnamefont {R.~M.}\ \bibnamefont
  {Stevenson}}, \bibinfo {author} {\bibfnamefont {A.~J.}\ \bibnamefont
  {Bennett}},\ and\ \bibinfo {author} {\bibfnamefont {A.~J.}\ \bibnamefont
  {Shields}},\ }\bibfield  {title} {\bibinfo {title} {{Electrically operated
  entangled light sources based on quantum dots}},\ }in\ \href@noop {} {\emph
  {\bibinfo {booktitle} {{Quantum dots}}}},\ \bibinfo {editor} {edited by\
  \bibinfo {editor} {\bibfnamefont {A.~G.}\ \bibnamefont {Tartakovskii}}}\
  (\bibinfo  {publisher} {{Cambridge University Press}},\ \bibinfo {address}
  {Cambridge and New York},\ \bibinfo {year} {2012})\ pp.\ \bibinfo {pages}
  {319--340}\BibitemShut {NoStop}%
\bibitem [{\citenamefont {Kantner}\ \emph {et~al.}(2017)\citenamefont
  {Kantner}, \citenamefont {Mittnenzweig},\ and\ \citenamefont
  {Koprucki}}]{Kantner.2017}%
  \BibitemOpen
  \bibfield  {author} {\bibinfo {author} {\bibfnamefont {M.}~\bibnamefont
  {Kantner}}, \bibinfo {author} {\bibfnamefont {M.}~\bibnamefont
  {Mittnenzweig}},\ and\ \bibinfo {author} {\bibfnamefont {T.}~\bibnamefont
  {Koprucki}},\ }\href {https://doi.org/10.1103/PhysRevB.96.205301} {\bibfield
  {journal} {\bibinfo  {journal} {Phys. Rev. B}\ }\textbf {\bibinfo {volume}
  {96}},\ \bibinfo {pages} {205301} (\bibinfo {year} {2017})}\BibitemShut
  {NoStop}%
\bibitem [{\citenamefont {Bennett}\ \emph {et~al.}(2005)\citenamefont
  {Bennett}, \citenamefont {Unitt}, \citenamefont {See}, \citenamefont
  {Shields}, \citenamefont {Atkinson}, \citenamefont {Cooper},\ and\
  \citenamefont {Ritchie}}]{Bennett.2005}%
  \BibitemOpen
  \bibfield  {author} {\bibinfo {author} {\bibfnamefont {A.~J.}\ \bibnamefont
  {Bennett}}, \bibinfo {author} {\bibfnamefont {D.~C.}\ \bibnamefont {Unitt}},
  \bibinfo {author} {\bibfnamefont {P.}~\bibnamefont {See}}, \bibinfo {author}
  {\bibfnamefont {A.~J.}\ \bibnamefont {Shields}}, \bibinfo {author}
  {\bibfnamefont {P.}~\bibnamefont {Atkinson}}, \bibinfo {author}
  {\bibfnamefont {K.}~\bibnamefont {Cooper}},\ and\ \bibinfo {author}
  {\bibfnamefont {D.~A.}\ \bibnamefont {Ritchie}},\ }\href
  {https://doi.org/10.1103/PhysRevB.72.033316} {\bibfield  {journal} {\bibinfo
  {journal} {Phys. Rev. B}\ }\textbf {\bibinfo {volume} {72}},\ \bibinfo
  {pages} {409} (\bibinfo {year} {2005})}\BibitemShut {NoStop}%
\bibitem [{\citenamefont {Hargart}\ \emph {et~al.}(2013)\citenamefont
  {Hargart}, \citenamefont {Kessler}, \citenamefont {Schwarzbäck},
  \citenamefont {Koroknay}, \citenamefont {Weidenfeld}, \citenamefont
  {Jetter},\ and\ \citenamefont {Michler}}]{Hargart.2013}%
  \BibitemOpen
  \bibfield  {author} {\bibinfo {author} {\bibfnamefont {F.}~\bibnamefont
  {Hargart}}, \bibinfo {author} {\bibfnamefont {C.~A.}\ \bibnamefont
  {Kessler}}, \bibinfo {author} {\bibfnamefont {T.}~\bibnamefont
  {Schwarzbäck}}, \bibinfo {author} {\bibfnamefont {E.}~\bibnamefont
  {Koroknay}}, \bibinfo {author} {\bibfnamefont {S.}~\bibnamefont
  {Weidenfeld}}, \bibinfo {author} {\bibfnamefont {M.}~\bibnamefont {Jetter}},\
  and\ \bibinfo {author} {\bibfnamefont {P.}~\bibnamefont {Michler}},\ }\href
  {https://doi.org/10.1063/1.4774392} {\bibfield  {journal} {\bibinfo
  {journal} {Appl. Phys. Lett.}\ }\textbf {\bibinfo {volume} {102}},\ \bibinfo
  {pages} {011126} (\bibinfo {year} {2013})}\BibitemShut {NoStop}%
\bibitem [{\citenamefont {{International Telecommunication
  Union}}(2015)}]{InternationalTelecommunicationUnion.2015}%
  \BibitemOpen
  \bibfield  {author} {\bibinfo {author} {\bibnamefont {{International
  Telecommunication Union}}},\ }\href@noop {} {\bibinfo {title} {{Nomenclature
  of the frequency and wavelengh bands used in telecommunications:
  Recommendation ITU-R V.431-8}}} (\bibinfo {year} {2015})\BibitemShut
  {NoStop}%
\bibitem [{\citenamefont {Ward}\ \emph {et~al.}(2014)\citenamefont {Ward},
  \citenamefont {Dean}, \citenamefont {Stevenson}, \citenamefont {Bennett},
  \citenamefont {Ellis}, \citenamefont {Cooper}, \citenamefont {Farrer},
  \citenamefont {Nicoll}, \citenamefont {Ritchie},\ and\ \citenamefont
  {Shields}}]{Ward.2014}%
  \BibitemOpen
  \bibfield  {author} {\bibinfo {author} {\bibfnamefont {M.~B.}\ \bibnamefont
  {Ward}}, \bibinfo {author} {\bibfnamefont {M.~C.}\ \bibnamefont {Dean}},
  \bibinfo {author} {\bibfnamefont {R.~M.}\ \bibnamefont {Stevenson}}, \bibinfo
  {author} {\bibfnamefont {A.~J.}\ \bibnamefont {Bennett}}, \bibinfo {author}
  {\bibfnamefont {D.~J.~P.}\ \bibnamefont {Ellis}}, \bibinfo {author}
  {\bibfnamefont {K.}~\bibnamefont {Cooper}}, \bibinfo {author} {\bibfnamefont
  {I.}~\bibnamefont {Farrer}}, \bibinfo {author} {\bibfnamefont {C.~A.}\
  \bibnamefont {Nicoll}}, \bibinfo {author} {\bibfnamefont {D.~A.}\
  \bibnamefont {Ritchie}},\ and\ \bibinfo {author} {\bibfnamefont {A.~J.}\
  \bibnamefont {Shields}},\ }\href {https://doi.org/10.1038/ncomms4316}
  {\bibfield  {journal} {\bibinfo  {journal} {Nat Commun}\ }\textbf {\bibinfo
  {volume} {5}},\ \bibinfo {pages} {3316} (\bibinfo {year} {2014})}\BibitemShut
  {NoStop}%
\bibitem [{\citenamefont {Scarani}\ \emph {et~al.}(2009)\citenamefont
  {Scarani}, \citenamefont {Bechmann-Pasquinucci}, \citenamefont {Cerf},
  \citenamefont {Dušek}, \citenamefont {Lütkenhaus},\ and\ \citenamefont
  {Peev}}]{Scarani.2009}%
  \BibitemOpen
  \bibfield  {author} {\bibinfo {author} {\bibfnamefont {V.}~\bibnamefont
  {Scarani}}, \bibinfo {author} {\bibfnamefont {H.}~\bibnamefont
  {Bechmann-Pasquinucci}}, \bibinfo {author} {\bibfnamefont {N.~J.}\
  \bibnamefont {Cerf}}, \bibinfo {author} {\bibfnamefont {M.}~\bibnamefont
  {Dušek}}, \bibinfo {author} {\bibfnamefont {N.}~\bibnamefont
  {Lütkenhaus}},\ and\ \bibinfo {author} {\bibfnamefont {M.}~\bibnamefont
  {Peev}},\ }\href {https://doi.org/10.1103/RevModPhys.81.1301} {\bibfield
  {journal} {\bibinfo  {journal} {Rev. Mod. Phys.}\ }\textbf {\bibinfo {volume}
  {81}},\ \bibinfo {pages} {1301} (\bibinfo {year} {2009})}\BibitemShut
  {NoStop}%
\bibitem [{\citenamefont {Chau}(2002)}]{Chau.2002}%
  \BibitemOpen
  \bibfield  {author} {\bibinfo {author} {\bibfnamefont {H.~F.}\ \bibnamefont
  {Chau}},\ }\href {https://doi.org/10.1103/PhysRevA.66.060302} {\bibfield
  {journal} {\bibinfo  {journal} {Phys. Rev. A}\ }\textbf {\bibinfo {volume}
  {66}},\ \bibinfo {pages} {802} (\bibinfo {year} {2002})}\BibitemShut
  {NoStop}%
\bibitem [{\citenamefont {Dousse}\ \emph {et~al.}(2010)\citenamefont {Dousse},
  \citenamefont {Suffczyński}, \citenamefont {Beveratos}, \citenamefont
  {Krebs}, \citenamefont {Lemaître}, \citenamefont {Sagnes}, \citenamefont
  {Bloch}, \citenamefont {Voisin},\ and\ \citenamefont
  {Senellart}}]{Dousse.2010}%
  \BibitemOpen
  \bibfield  {author} {\bibinfo {author} {\bibfnamefont {A.}~\bibnamefont
  {Dousse}}, \bibinfo {author} {\bibfnamefont {J.}~\bibnamefont
  {Suffczyński}}, \bibinfo {author} {\bibfnamefont {A.}~\bibnamefont
  {Beveratos}}, \bibinfo {author} {\bibfnamefont {O.}~\bibnamefont {Krebs}},
  \bibinfo {author} {\bibfnamefont {A.}~\bibnamefont {Lemaître}}, \bibinfo
  {author} {\bibfnamefont {I.}~\bibnamefont {Sagnes}}, \bibinfo {author}
  {\bibfnamefont {J.}~\bibnamefont {Bloch}}, \bibinfo {author} {\bibfnamefont
  {P.}~\bibnamefont {Voisin}},\ and\ \bibinfo {author} {\bibfnamefont
  {P.}~\bibnamefont {Senellart}},\ }\href {https://doi.org/10.1038/nature09148}
  {\bibfield  {journal} {\bibinfo  {journal} {Nature}\ }\textbf {\bibinfo
  {volume} {466}},\ \bibinfo {pages} {217} (\bibinfo {year}
  {2010})}\BibitemShut {NoStop}%
\bibitem [{\citenamefont {Arcari}\ \emph {et~al.}(2014)\citenamefont {Arcari},
  \citenamefont {Söllner}, \citenamefont {Javadi}, \citenamefont {{Lindskov
  Hansen}}, \citenamefont {Mahmoodian}, \citenamefont {Liu}, \citenamefont
  {Thyrrestrup}, \citenamefont {Lee}, \citenamefont {Song}, \citenamefont
  {Stobbe},\ and\ \citenamefont {Lodahl}}]{Arcari.2014}%
  \BibitemOpen
  \bibfield  {author} {\bibinfo {author} {\bibfnamefont {M.}~\bibnamefont
  {Arcari}}, \bibinfo {author} {\bibfnamefont {I.}~\bibnamefont {Söllner}},
  \bibinfo {author} {\bibfnamefont {A.}~\bibnamefont {Javadi}}, \bibinfo
  {author} {\bibfnamefont {S.}~\bibnamefont {{Lindskov Hansen}}}, \bibinfo
  {author} {\bibfnamefont {S.}~\bibnamefont {Mahmoodian}}, \bibinfo {author}
  {\bibfnamefont {J.}~\bibnamefont {Liu}}, \bibinfo {author} {\bibfnamefont
  {H.}~\bibnamefont {Thyrrestrup}}, \bibinfo {author} {\bibfnamefont {E.~H.}\
  \bibnamefont {Lee}}, \bibinfo {author} {\bibfnamefont {J.~D.}\ \bibnamefont
  {Song}}, \bibinfo {author} {\bibfnamefont {S.}~\bibnamefont {Stobbe}},\ and\
  \bibinfo {author} {\bibfnamefont {P.}~\bibnamefont {Lodahl}},\ }\href
  {https://doi.org/10.1103/PhysRevLett.113.093603} {\bibfield  {journal}
  {\bibinfo  {journal} {Phys. Rev. Lett.}\ }\textbf {\bibinfo {volume} {113}},\
  \bibinfo {pages} {093603} (\bibinfo {year} {2014})}\BibitemShut {NoStop}%
\end{thebibliography}%
%

\end{document}